\documentclass[]{iopart}

\usepackage{amssymb} 
\usepackage{color}
\usepackage{url}
\usepackage{psfrag}
\usepackage{natbib}
\usepackage[latin1]{inputenc}
\usepackage{amssymb}
\usepackage{geometry}  
\usepackage{graphicx}
\usepackage{dsfont}
\usepackage{txfonts}
\usepackage{subfigure}

\usepackage{aas_macros}
\def\newblock{\hskip .11em plus .33em minus .07em}

\begin{document}

\title{Gravitational waves from supernova matter}

\author{S. Scheidegger, S.C. Whitehouse, R. K\"appeli, and M. Liebend\"orfer}

\address{Institute of Physics, 
Basel University
Klingelbergstrasse 82
CH-4056 Basel , Switzerland}
\ead{simon.scheidegger@unibas.ch}

\begin{abstract}
We have performed a set of 11 three-dimensional
magnetohydrodynamical core-collapse supernova
simulations in order to investigate the dependencies of the
gravitational wave signal on the progenitor's initial conditions.
We study the effects of the initial central
angular velocity and different 
variants of neutrino transport.
Our models are started up from a 15$M_{\odot}$ progenitor and incorporate
an effective general relativistic gravitational potential 
and a finite temperature nuclear equation of state. 
Furthermore, the electron
flavour neutrino transport is tracked by
efficient algorithms for the radiative transfer 
of massless fermions. 
We find that non- and slowly rotating models 
show gravitational wave
emission due to prompt- and lepton driven convection that reveals
details about the hydrodynamical state of the 
fluid inside the protoneutron stars.
Furthermore we show that protoneutron stars can become
dynamically unstable to rotational 
instabilities at $T/|W|$ values as low
as $\sim2\%$ at core bounce.
We point out that the inclusion of deleptonization during the
postbounce phase is very important for the quantitative GW
prediction, as it enhances the absolute
values of the gravitational wave trains up to a factor of ten with
respect to a lepton-conserving treatment.
\end{abstract}

\pacs{04.30.Db, 95.30.Qd, 97.60.Bw}


\section{Introduction}
\label{section:intro}

Stars in the mass range 
$8M_{\odot}\lesssim M \lesssim 40M_{\odot}$
end their lives
in a core-collapse supernova (CCSN). 
However, at present the fundamental explosion mechanism, 
which causes a star to lose its envelope
by a yet uncertain combination of factors including
neutrino heating, rotation, hydrodynamical instabilities,
core g-mode oscillations and magnetic fields, is still under debate 
(for a review, see e.g. \cite{2007PhR...442...38J}).
As strong indications both from
theory and observations exist that 
CCSNe show aspherical, multidimensional features 
\cite{1994ApJ...435..339H,2006Natur.440..505L}, 
there is a reasonable hope
that a small amount of the released 
binding energy will also be emitted as
gravitational waves (GWs), thus delivering 
us first-hand information about
the dynamics and the state of matter
at the centre of the star. 
GW emission from CCSNe
were suggested to arise from i) 
axisymmetric rotational core collapse and bounce 
\cite{1997A&A...320..209Z,2004PhRvD..69l4004K,
2007PhRvL..98y1101D,2008PhRvD..78f4056D}
ii) prompt-, neutrino-driven postbounce convection 
and anisotropic neutrino emission
\cite{1997A&A...317..140M,M2004,2007ApJ...655..406K,2009A&A...496..475M,
2009ApJ...707.1173M,2009ApJ...704..951K,2009ApJ...697L.133K}, 
iii) protoneutron star (PNS) g-mode oscillations
\cite{2006PhRvL..96t1102O}
and iv) nonaxisymmetric rotational instabilities 
\cite{2003ApJ...595..352S,2005ApJ...618L..37W,
2005ApJ...625L.119O,
2006ApJ...651.1068O,2007CQGra..24..139O,2007PhRvL..98z1101O,
2008A&A...490..231S,2010arXiv1001.1570S}
For recent reviews with a more complete 
list of references, 
see \cite{2006RPPh...69..971K,2009CQGra..26f3001O}.
However, only i) can be considered as being
well understood as far as the physics of the collapse
is concerned, since only theses models 
incorporate all
relevant input physics known at present
\cite{2008PhRvD..78f4056D}
(there are, though, still large uncertainties with 
respect to the progenitor star, e.g. rotation
profiles, magnetic fields, and inhomogenities
from convection).
The prediction of all other suggested 
emission scenarios (ii-iv) still neglect, 
to a certain extent, dominant physics features
due to the diversity and complexity
of the CCSN problem on the one hand side
and restrictions of available computer 
power on the other side.
Hence, the computational resources were 
so far either spent on highly 
accurate neutrino transport 
(e.g. \cite{2004ApJS..150..263L,2008ApJ...685.1069O,2009A&A...496..475M}) 
while neglecting other physical 
degrees of freedom such as 
magnetic fields, or 
focus on a general relativistic treatment and/or 
3D fluid effects such as 
accretion funnels, rotation rate and convection,
but approximate or even 
neglect the important micro physics.
Only recently have detailed 3D computer models 
of CCSN become
feasible with the emerging
power of tens of thousands CPUs unified
in a single supercomputer.
Such detailed simulations are absolutely indispensable 
for the following reasons: a) GW astronomy 
requires not only
very sensitive detectors, but also
depends on extensive data processing
of the detector output on the basis
of reliable GW estimates \cite{abbott}.
b) The temperatures and
densities inside a supernova core 
exceed the range that is easily accessible by terrestrial
experiments. Thus, it will be impossible for the foreseeable 
future to construct a unique finite temperature
equation of state (EoS) for hot and dense
matter based on experimentally verified data.
Therefore, models with different parameter settings must be
run and their computed wave form output then can
be compared with actual detector data.
Hence, modelling will bridge 
the gap between theory and measurement and 
allowing the use of use of CCSNe
as laboratory for exotic nuclear and particle physics 
\cite{2009NuPhA.827..573L}.
In this paper, we will present the GW 
analysis of a 
set of 11 three-dimensional
ideal magnetohydrodynamical (MHD) core-collapse simulations.
We will focus our study on the imprint 
of 3D nonaxisymmetric features onto the GW signature. 
Our calculations include presupernova
models from stellar evolution calculations,
a finite-temperature nuclear EoS and 
a computationally efficient treatment 
of deleptonization during the collapse phase.
General relativistic corrections to the 
spherically symmetric Newtonian gravitational
potential are taken into account.
Moreover, while several models 
incorporate 
long-term neutrino physics by means
of a leakage scheme, we also present
the first results of a model
which includes a neutrino
transport approximation in the postbounce
phase that takes
into account both neutrino heating and
cooling.
As for the progenitor, we systematically
investigate the effects of the spatial grid
resolution, the neutrino transport physics and
the precollapse rotation rate  
with respect 
to its influence on the nonaxisymmetric matter dynamics.

This paper is organised as follows. In section 2 we briefly 
describe the initial model configurations and the 
numerical techniques employed for their temporal 
evolution. Furthermore, we review the tools 
used for the GW and data analysis.
Section 3 collects the results of our simulations.
Finally, section 4 contains our conclusions and an outlook of
our future research.


\subsection{Description of the magnetohydrodynamical models}
\label{section:MHD}

For the 3D Newtonian ideal MHD  
CCSN simulations presented in this paper, 
we use the \texttt{FISH} code \cite{2009arXiv0910.2854K}.
The gravitational potential is calculated via  
a spherically symmetric mass integration that includes 
radial general relativistic corrections \cite{Marek2006}.
The 3D computational domain consists of a central cube of either 
$600^3$ or $1000^3$ cells,
treated in equidistant Cartesian coordinates with a
grid spacing of 1km or 0.6km. It is, as explained 
in \cite{2008A&A...490..231S}, 
embedded in a larger spherically symmetric
computational domain that is treated by the time-implicit 
hydrodynamics code `Agile' \cite{Liebend2002}.
Closure for the MHD equations is obtained 
by the softest version of the finite-temperature nuclear EoS of 
\cite{Lattimer1991}. 
The inclusion of neutrino physics is
an essential ingredient of CCSNe simulations, as 
$\sim$99\% of the released binding energy is converted 
into neutrinos of all flavours. 
Their complex interactions with matter 
(e.g. \cite{2007PhR...442...38J}) are believed to drive the
supernova explosion dynamics in the outer layers as well as
deleponizing the PNS to its compact final stage as a neutron star.
As the Boltzmann neutrino transport equation can only
be numerically solved in a complete form 
in spherical symmetry on today's supercomputers \cite{Mezzacappa2005},
our 3D simulations must rely on several feasible 
approximations which capture the
dominant features of the neutrino physics.
As for the treatment of the deleptonization during the
collapse phase, we apply a simple and computationally efficient
$Y_{e}$ vs. $\rho$ parametrization
scheme which is based on data from detailed 
1D radiation-hydrodynamics calculations \cite{Liebendorfer2005}.
For this we use the results obtained with the  \texttt{Agile-Boltztran} 
code \cite{Liebend2005}, including the above-mentioned EoS 
and the electron capture rates 
from \cite{1985ApJS...58..771B}.
Around core bounce, this scheme breaks down as it cannot 
model the neutronization burst.
After core bounce, the neutrino transport thus is 
tracked for several models 
via a partial (i.e. a leakage scheme) or full implementation of 
the isotropic diffusion source approximation scheme
(IDSA, \cite{2009ApJ...698.1174L}).
The IDSA decomposes the distribution function
$f$ of neutrinos into two components, a trapped
component $f^{t}$ and a streaming component $f^{s}$, representing
neutrinos of a given species and energy which 
find the local zone opaque or transparent, respectively.
The total distribution function is the sum of the two
components, $f=f^{t} + f^{s}$. The two components
are evolved using separate numerical techniques, coupled
by a diffusion source term $\Sigma$.
The source term $\Sigma$ converts trapped into streaming
particles and vice versa. We determine it from the requirement
that the temporal change of $f^{t}$ 
has to reproduce the diffusion limit in the limit of small mean 
free path.
Note that our leakage scheme significantly overestimates the
deleptonization in and around the neutrinosphere region,
as it neglects any absorption 
of transported neutrinos by discarding the streaming 
component ($f^{s}=0$). 
\begin{table}
\caption{\label{tabone} Summary of the models' initial 
conditions and GW related quantities. $\Omega_{c,i}$ [rads$^{-1}$] is
the precollapse central angular velocity, while $\beta = T/|W|$ is 
the ratio of rotational to gravitational energy. $\rho_{c,b}$ 
[$10^{14}gcm^{-3}$] is the maximum central density 
at the time of core bounce.
E$_{GW}$ [$10^{-9}M_{\odot}c^2$] is the energy emitted as GWs.
$f_{b}$ [Hz] denotes the peak frequency of the GW burst at bounce, while 
$f_{TW}$ [Hz] stands for the spectral peak from the narrow band emission
caused by a low $T/|W|$ instability. $t_{f}$ [ms] is the
time after core bounce when the simulation was stopped.
}

\begin{indented}
\lineup
\label{tab1}
\item[]\begin{tabular}{@{}*{8}{l}}
\br
$\0\0Model$ & $\Omega_{c,i}$ & $\beta_{i}$ & $\beta_{b}$ &$\rho_{c,b}$ &$\0f_{TW}$&$E_{GW}$ & $t_{f}$\cr
\mr
\0\0 R0           & 0    & 0                 & 0                    & 4.39 & - & 0.02 &130 \cr
\0\0 R0$_{IDSA}$  & 0    & 0                 & 0                    & 4.34 & - & 0.01 & 81 \cr

\0\0 R1$_{HR}$    & 0.3  & 0.6$\cdot10^{-5}$ & 1.7$\cdot10^{-4}$    & 4.36 & - & 0.24 & 25\cr
\0\0 R1$_{L}$     & 0.3  & 0.6$\cdot10^{-5}$ & 1.8$\cdot10^{-4}$    & 4.38 & - & 0.10  & 93\cr

\mr

\0\0 R2          & 3.14 & 0.6$\cdot10^{-3}$ & $1.6\cdot10^{-2}$     & 4.27 & - & 5.5 &127\cr

\0\0 R3             & 3.93 & 1.0$\cdot10^{-3}$ & $2.3\cdot10^{-2}$  & 4.16 & 670 & 14&106\cr

\0\0 R4            & 4.71 & 1.4$\cdot10^{-3}$ & $3.2\cdot10^{-2}$  & 4.04  & 615 & 35&64 \cr

\0\0 R5           & 6.28 & 2.6$\cdot10^{-3}$ & $5.2\cdot10^{-2}$    & 3.80 & 725 & 59&63\cr
\0\0 R5$_{L}$     & 6.28 & 2.6$\cdot10^{-3}$ & $5.1\cdot10^{-2}$    & 3.65 & 909 &214&197\cr

\0\0 R6           & 9.42 & 5.7$\cdot10^{-3}$ & $8.6\cdot10^{-2}$   & 3.22  & 662 & 77&99\cr
\0\0 R7          & 12.57 & 1.0$\cdot10^{-2}$ & $10.2\cdot10^{-2}$   & 2.47 & 727& 12&93\cr

\br
\end{tabular}
\end{indented}
\end{table}
The presupernova stellar models stem from 
Newtonian 1D stellar evolution calculations and hence 
may not cover all possible  
states prior to the collapse of a multidimensional star.
Therefore we construct the initial conditions 
of our simulations by
a parametric approach.
We employ a solar-metallicity 15$M_{\odot}$
progenitor of \cite{Woosley1995}, and set it into rotation
according to a shell-type 
rotation law of \cite{1985A&A...146..260E} with a
shellular quadratic cutoff at 500km radius.
The initial magnitude of the magnetic field strength for all
models is fixed at values suggested by \cite{2005ApJ...626..350H}.

\subsection{Gravitational Wave extraction}
\label{section:GW}

We employ the Newtonian quadrupole formula in the 
\textit{first-moment of momentum density formulation} 
\cite{1990ApJ...351..588F}
to extract the GWs from our simulation data.
Note that the quadrupole formula is not gauge invariant
and only valid in the Newtonian slow-motion limit 
\cite{1973grav.book.....M}. Nevertheless, 
it was shown by \cite{2003PhRvD..68j4020S} in comparative
tests to work sufficiently well
compared to more sophisticated methods, 
as it preserves phase while being off 
in amplitude by $\sim$10\%.


\section{Results}
\label{section:results}

\subsection{Non- or slowly rotating core collapse}
\begin{figure}
\centering
\subfigure{\includegraphics[scale=0.47]{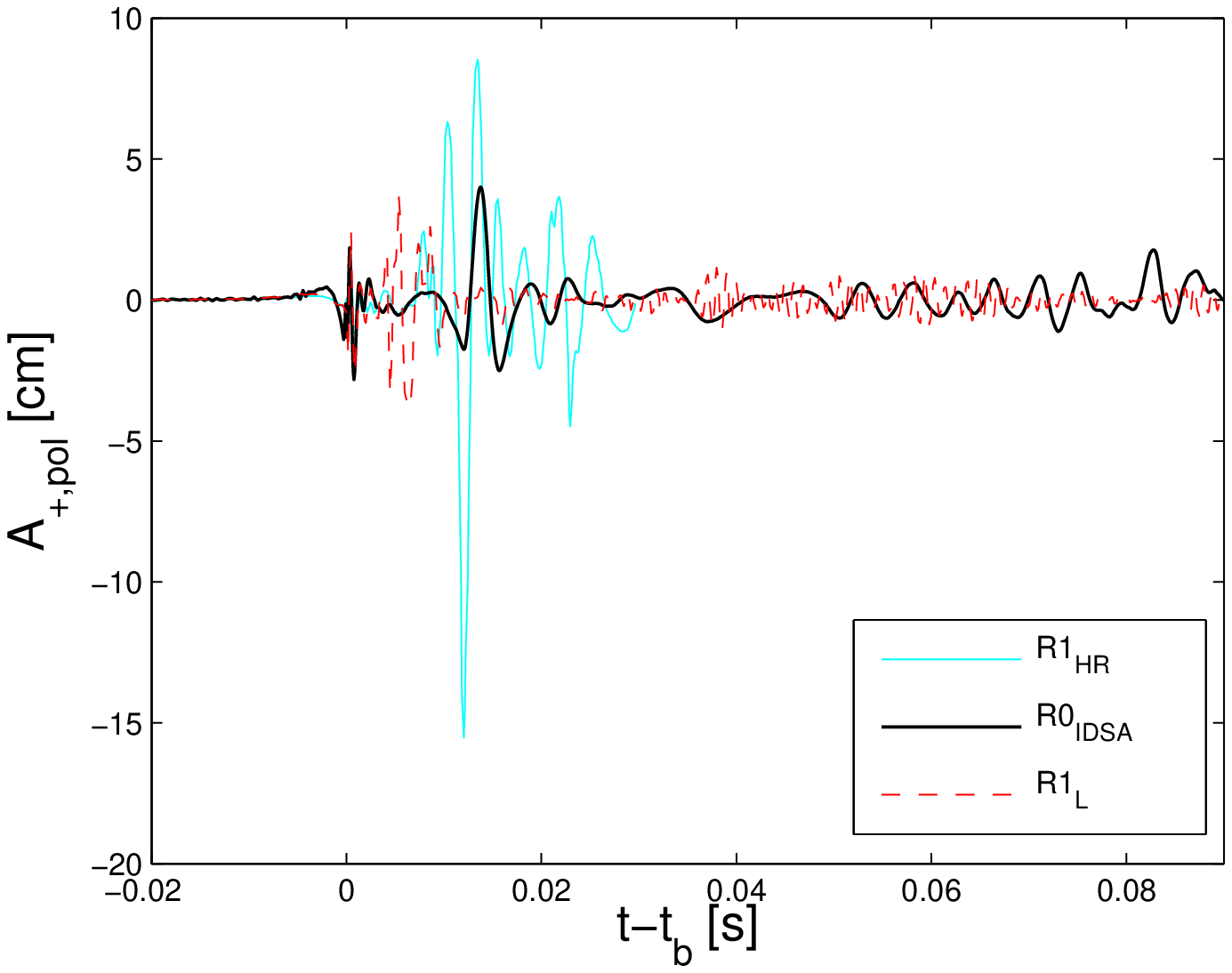}\label{fig:1a}}
\subfigure{\includegraphics[scale=0.47]{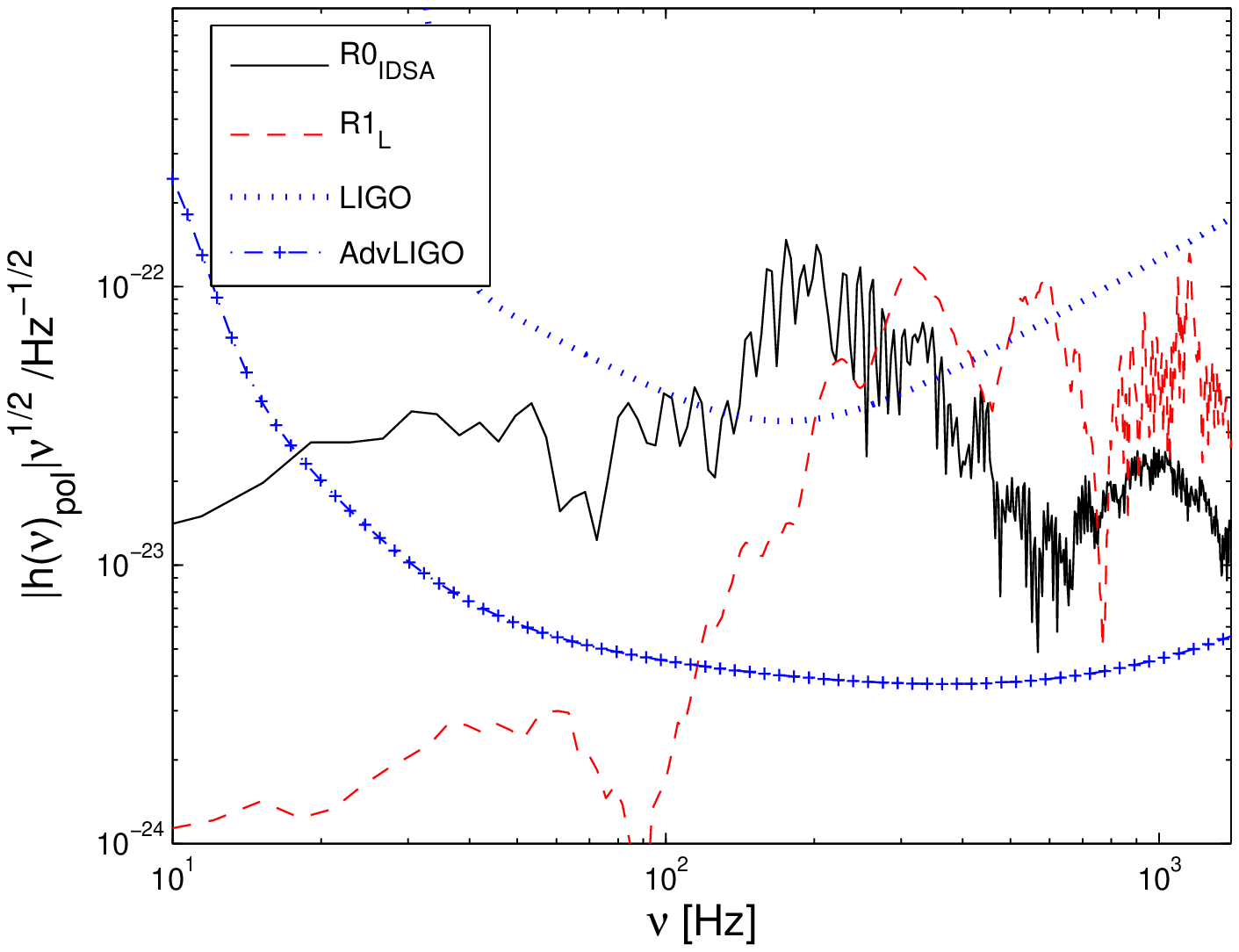}\label{fig:1b}}
\caption{\textbf{Left:} Time evolution of the 
GW $+$ polarization for a spectator located at the polar axis 
(Models R1$_{HR}$, R0$_{IDSA}$ and R1$_{L}$)
\textbf{Right:} 
Corresponding spectral energy distribution of models 
R0$_{IDSA}$ and R1$_{L}$ at a distance of 10kpc compared
with the LIGO strain sensitivity and the planned performance of
Advanced LIGO \cite{Shoemaker:2007}. Optimal orientation between source
and detector is assumed.}
\end{figure}
\begin{figure}
\centering
\subfigure{\includegraphics[width=7cm]{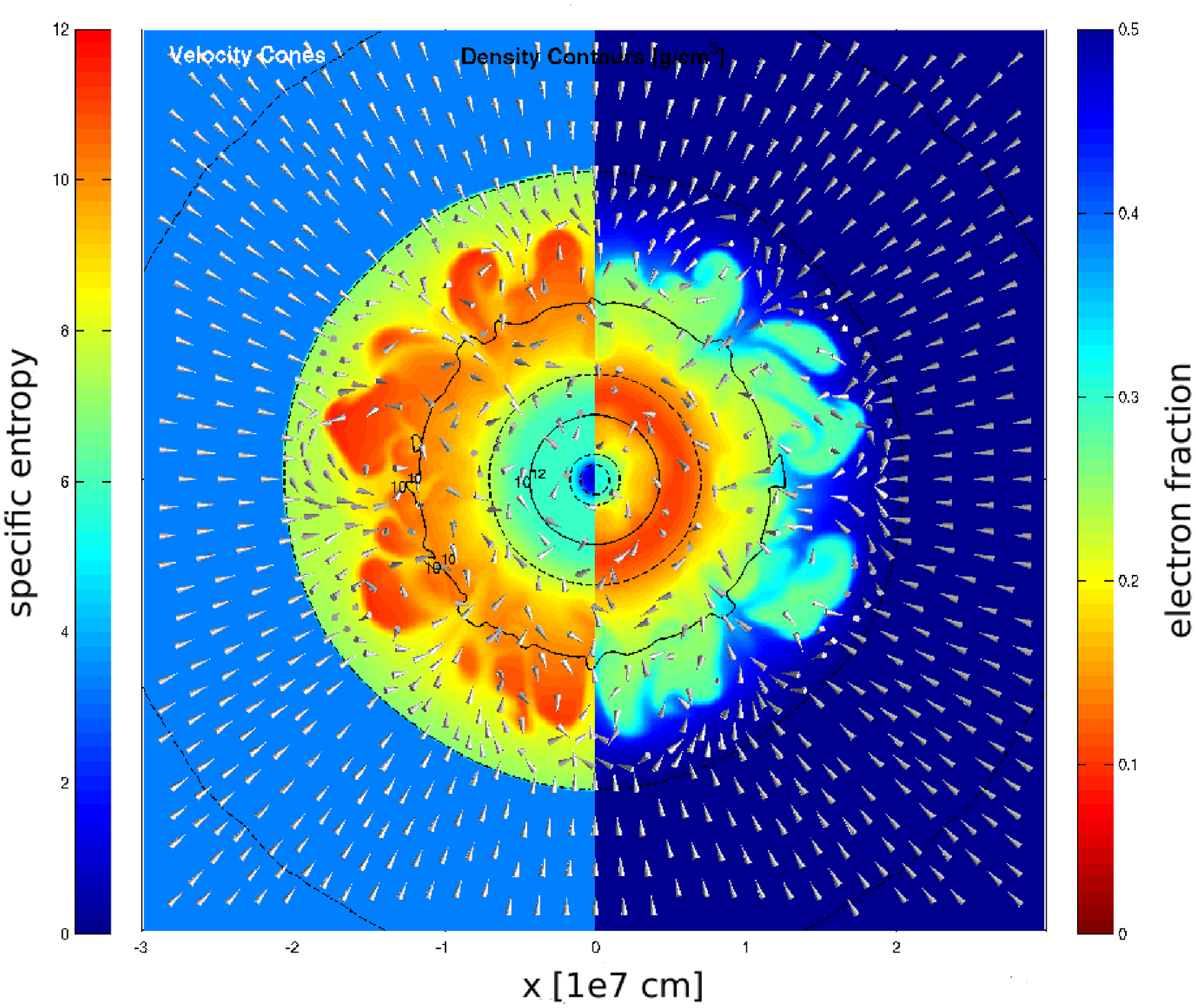}\label{fig:2a}}
\subfigure{\includegraphics[width=7.7cm]{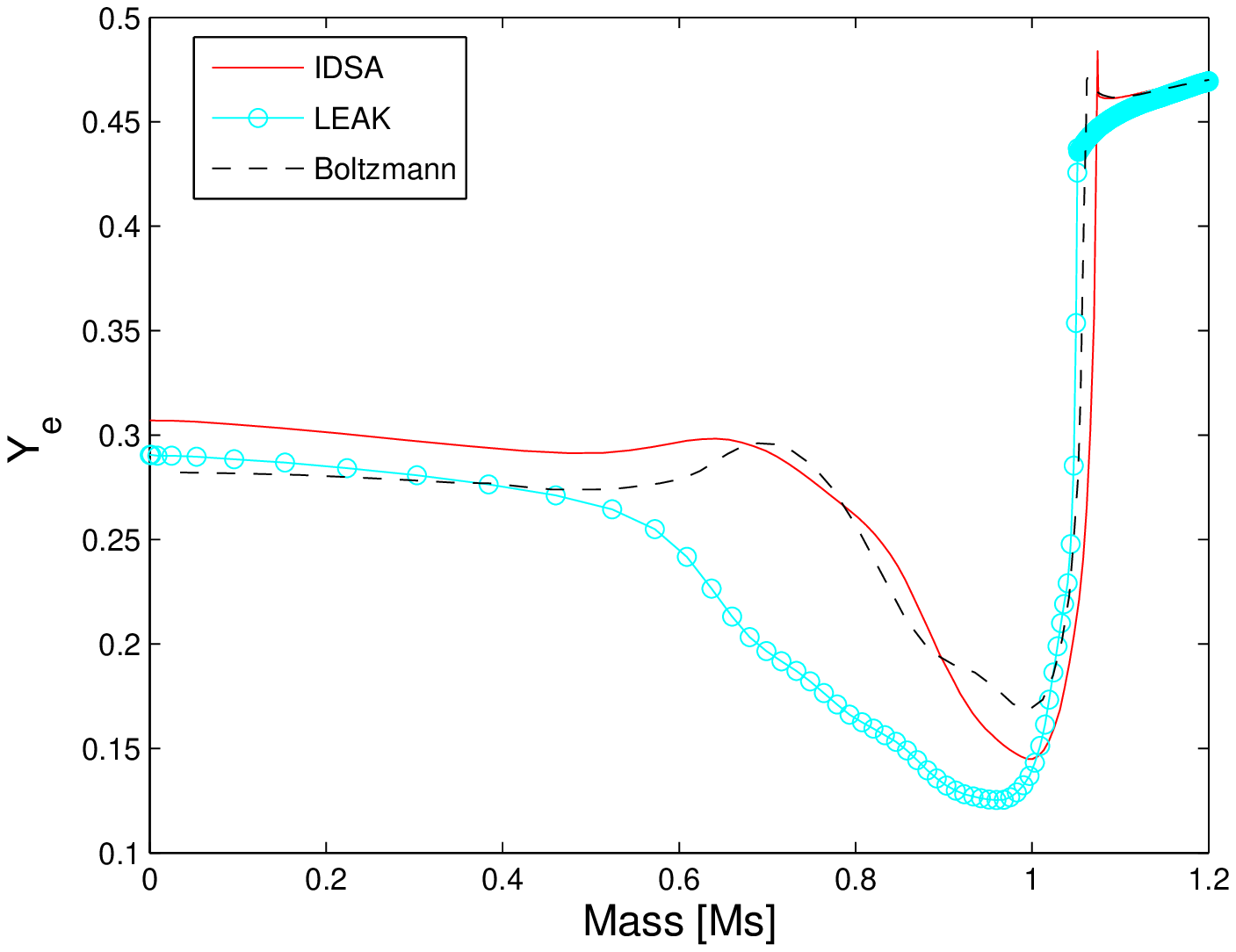}\label{fig:2b}}
\caption{\textbf{Left:} Model R0$_{IDSA}$'s 
specific entropy distribution [k$_{B}/$baryon] (left side)
and electron fraction $Y_{e}$ (right side) 50ms after core bounce. 
The innermost 600$^2$km$^2$ in the x-y plane
are displayed. The entropy color bar scales from 0 (blue) to 12 (red).
The $Y_{e}$ color bar accounts for values from 0 (red) to 0.5 (blue) 
\textbf{Right:} Comparison of the spherically averaged 
$Y_{e}$ profiles of models R0$_{L}$ (circled line, $`LEAK'$), 
R0$_{IDSA}$ (solid line, $`IDSA'$) with the spherically 
symmetric model G15 (dashed line, $`Boltzmann'$) as a
function of the enclosed mass at 5ms after bounce.
Model G15 is based on 
general relativistic
three-flavour neutrino transport \cite{Liebend2005}. 
}
\end{figure}
In order to study the influence of 
neutrino transport 
on the stochastic matter dynamics
in the early supernova stages ($t\lesssim100$ms after bounce),
without having other different physical parameters 
interfering, we carried out three simulations:
R0 (purely hydrodynamical postbounce evolution), 
R1$_{L}$ (includes a leakage scheme)
and R0$_{IDSA}$ (incorporates both neutrino cooling and heating).
Non- and slowly rotating progenitors 
($\Omega_{c,i} \leq 0.3$rads$^{-1}$ in our model set) 
all undergo quasi-spherically symmetric core collapse.
As the emission of GWs intrinsically depends on 
dynamical processes that deviate from spherical symmetry,
the collapse phase therefore does not provide any
kind of signal, as shown in Fig.\ref{fig:1a} for $t-t_{b}<0$. 
However, subsequent pressure-dominated core bounce, 
where the collapse is halted
due to the stiffening of the EoS at nuclear density
$\rho_{nuc}\approx 2\times 10^{14}$gcm$^{-3}$, 
launches a shock wave that plows through
the infalling layers, leaving behind a negative entropy
gradient.
Moreover, as soon as the shock 
breaks through the neutrino sphere $\sim5$ms after bounce, 
the immediate burst 
of electron neutrinos causes a negative 
lepton gradient at the edge of the PNS.
The combination of these two gradients 
form a convectively unstable region according 
to the Schwarzschild-Ledoux criterion 
\cite{1959flme.book.....L,1988PhR...163...63W}, 
which in turn induces a GW burst due to this
so-called `prompt' convection.
A detailed comparison of the models 
R0, R1$_{L}$, R0$_{IDSA}$ shows that 
all of them follow a similar dynamical behaviour
until about 20ms after bounce. At this stage, aspherities
leading to GW emission
are predominantly 
driven by the negative entropy gradient and not by the lepton
gradient.
Hence, the wave trains of all three models,
which are based on stochastic processes, 
fit each other relatively well 
both in amplitude (several cm) and spectra ($\sim150 - 500$Hz).
However, `prompt' convection depends, as it was
pointed out by \cite{2009CQGra..26f3001O},
not on the negative entropy gradient alone, but
also on numerical seed perturbations which are
introduced by the choice of the computational grid.
Hence, in order to test the dependence of 
our findings on the spatial resolution, 
we carried out model  R1$_{HR}$.
This better resolved simulation shows 
considerably smaller seed perturbations around $t-t_{b}\sim$ 0,
as grid alignment effects are better 
suppressed at core bounce; 
hence prompt convection
then is much weaker and a smaller GW amplitude ($\sim50\%$)
is emitted, as shown in Fig.\ref{fig:1a}.
However, better numerical resolution also leads 
to less numerical dissipation in the system, 
which eases the dynamical effects that follow. 
Thus, we find for $\sim10$ms $\lesssim t \lesssim20$ms
considerably stronger GW emission in R1$_{HR}$ 
compared to the 1km resolved
models, as indicated in Fig.\ref{fig:1a}.
The three representative simulation results
diverge strongly in the later postbounce evolution 
($t\gtrsim$ 20ms).
Convective overturn 
causes a smoothing of the entropy 
gradient. As a result, the GW amplitude in 
the hydrodynamical model R0 quickly decays
($t\lesssim30$ms after bounce) 
and is not revived
during the later evolution.
On the other hand, the negative radial lepton gradient
(see Figs.\ref{fig:2a} and \ref{fig:2b}) 
which is caused by the neutronization burst and the 
subsequent deleptonization, which we model only in 
R1$_{L}$ and R0$_{IDSA}$, now starts to drive 
convection inside the PNS. 
For the latter models, the so-called PNS convection 
\cite{2006ApJ...645..534D} 
exhibits similar maximum amplitudes of $\sim$1-2cm 
(Fig.\ref{fig:1a}), while differing from each other
strongly in the corresponding spectra, 
as displayed in Fig.\ref{fig:1b}. 
R1$_{L}$'s spectrum peaks between $\sim$600 -1000Hz,
while R0$_{IDSA}$'s frequency band peaks at 
values as low as $\sim$100Hz.
This affects the total energy $E_{GW}$ emitted
($\mathcal{O}(10^{-10})M_{\odot}c^2$ vs. $\mathcal{O}(10^{-11})M_{\odot}c^2$,
see Tab.\ref{tab1}), being one order
of magnitude higher for R1$_{L}$ due to
$dE_{GW}/df\propto f^2$.
We found the key controlling factor of this behaviour 
to be the radial location of the convectively
unstable zones and the related dynamical characteristical 
timescales $t_{dyn}$ involved. 
If we use as rough estimate t$_{dyn} \sim \Delta_{r}/\overline{c_{s}}$, 
\footnote{$\overline{c_{s}}=1/\Delta_{r}\int_{r}c_{s}(r)dr$ is
the radially averaged sound speed of a convectively unstable layer 
with a radial extension of $\Delta_{r}$.},
and apply typical values for the models R0$_{IDSA}$ and R1$_{L}$, 
we confirm the obtained values.
Furthermore, our leakage scheme significantly overestimates 
neutrino cooling processes, 
as one can see in Fig.\ref{fig:2b}.
There, the convectivly unstable layer is extended to radii
above nuclear densities, where matter still is opaque
for neutrinos and where the local
speed of sound assumes values far larger than in the 
case of model R0$_{IDSA}$. Hence, the dynamical timescale 
of R1$_{L}$ is considerably shorter and the 
spectral distribution is peaked at
higher values. 
When comparing the results found for model
R0$_{IDSA}$ with a 
very recent 2D study of \cite{2009A&A...496..475M}, 
where they carried out one simulation (cf. their model 
M15LS-2D) 
with comparable input physics (same $15M_{\odot}$ progenitor; 
same underlying finite-temperature EoS) and a very
sophisticated neutrino transport scheme,
we find very good agreement both in the 
amplitudes and frequencies.
Hence we conclude that 
the primary ingredient for supernova simulations which 
attempt a quantitative prediction of GWs from `prompt' 
and early PNS convection ($t\lesssim100$ms after bounce)
is the accurate radial location and size of convectively 
unstable layers. It defines the dynamical
behaviour and timescale of overturning matter
in this early supernova stage.

\subsection{Rotational core collapse \& nonaxisymmetric instability at low $T/|W|$}

\begin{figure}
\centering
\subfigure{\includegraphics[width=6cm,height=5.5cm]{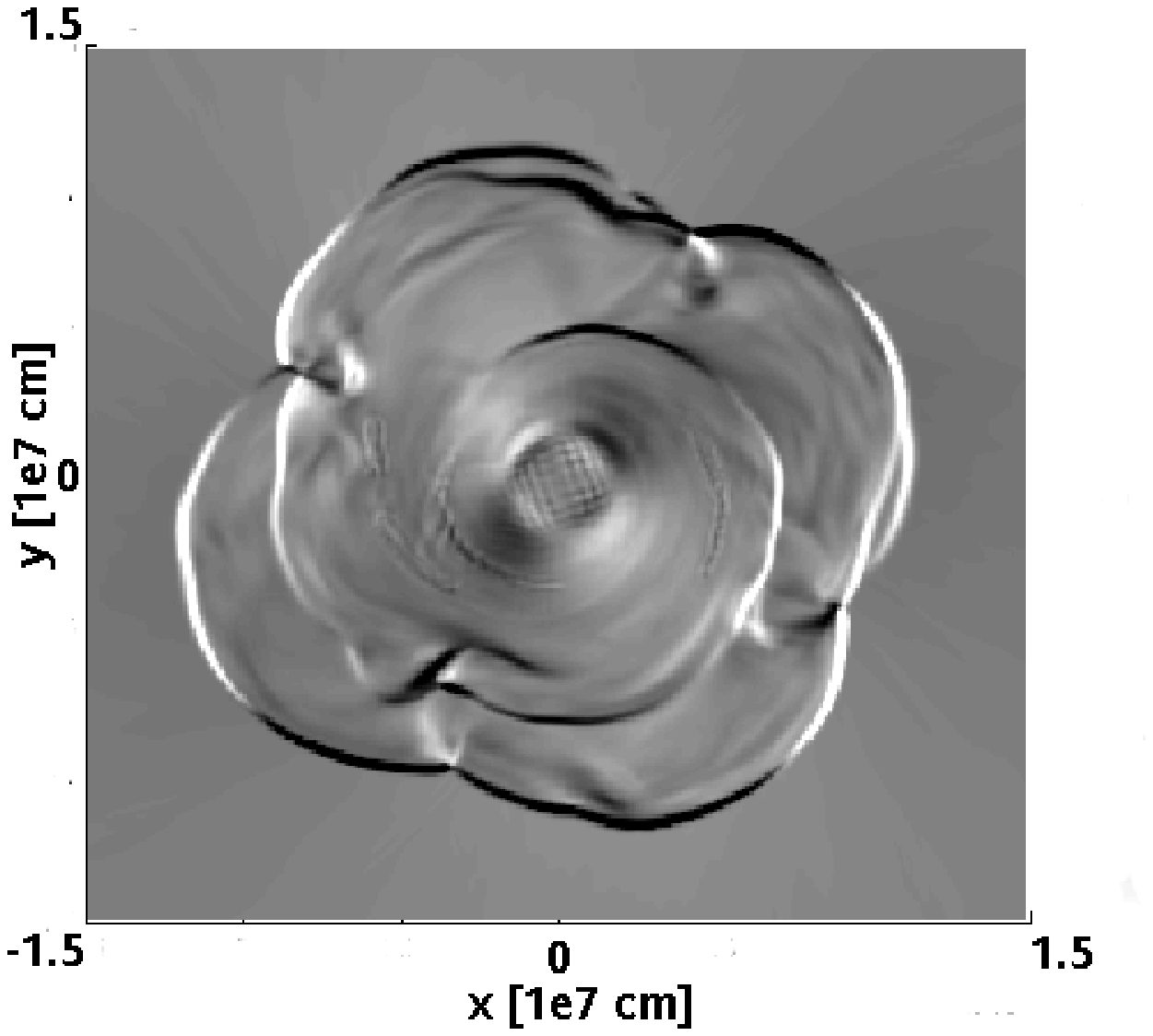}\label{fig:3a}}
\subfigure{\includegraphics[width=9.0cm,height=5.5cm]{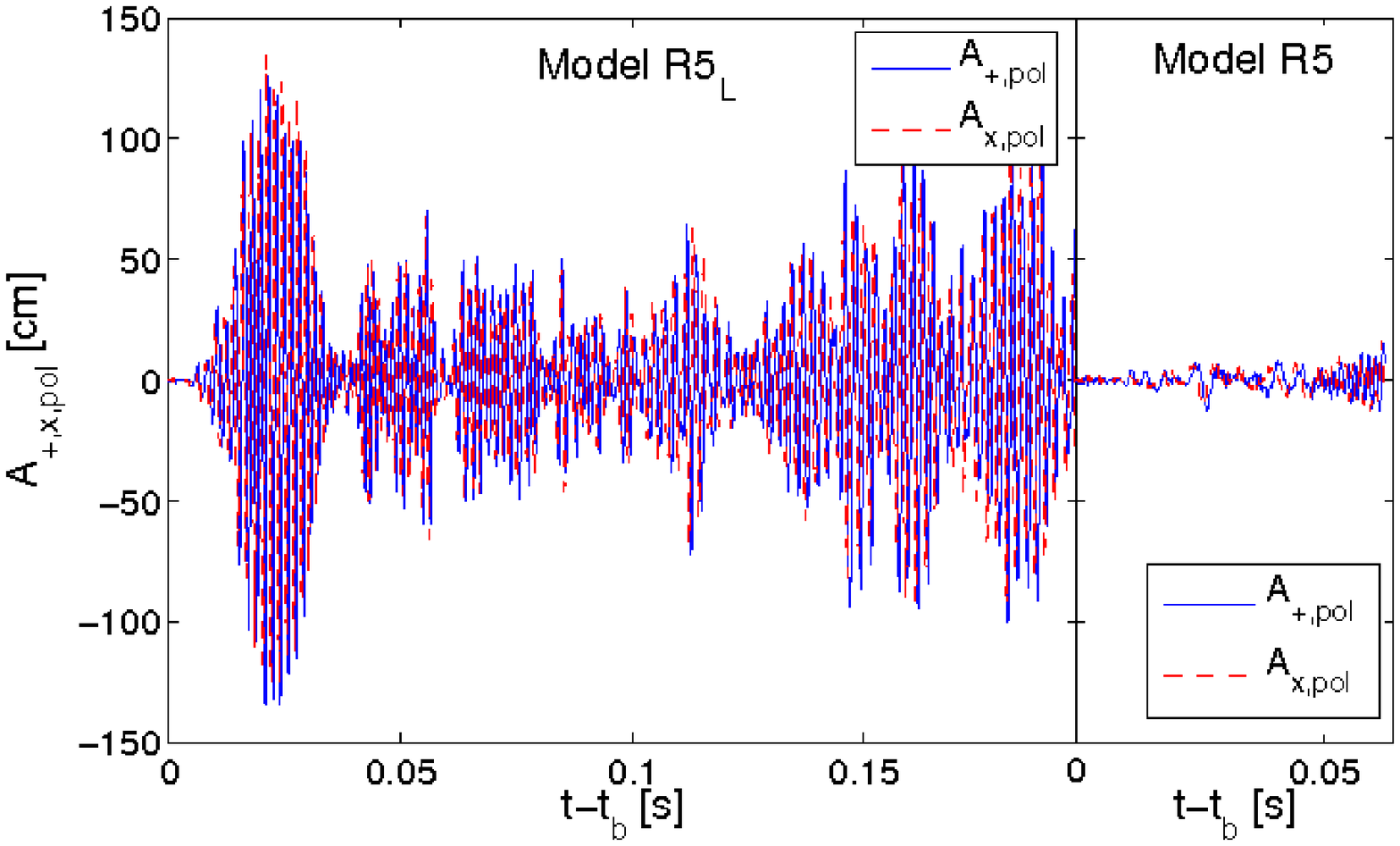}\label{fig:3b}}
\caption{\textbf{Left:} Vorticity's z-component 
w$_{z}=(\nabla\times v)_{z}$ of model R5$_{L}$ 
in the equatorial plane, 29ms after bounce, 
showing a dominant m=1 mode. 
The innermost 300$^2$km$^2$ are displayed, 
and the color is encoded in units of [s$^{-1}$], ranging
from -5000 (white) to 5000 (black)
\textbf{Right:} GW polarizations $+$ and $\times$ 
for models R5$_{L}$ and R5 as 
seen from an observer along the polar axis. Strong nonaxisymmetric
dynamics with m=2 components develop right after core bounce.
Hence, the two polarizations are shifted by a quarter
cycle, as one could expect from GWs emitted by a spinning bar.}
\end{figure}
Recently it has been argued based on numerical simulations 
of equilibrium neutron star models or full core-collapse simulations 
that differentially rotating PNS can be subject to 
non-axiymmetric rotational instabilities (see Fig.\ref{fig:3a}) 
at $\beta$ values ($\hat{=}T/|W|$, 
the ratio of rotational to gravitational energy) far below 
the ones known from the classical dynamical bar mode instability with a
threshold of $\beta_{dyn} = 27\%$, or the secular instability, which
is triggered at $\beta_{sec}\sim$ 14\% \cite{1978trs..book.....T},
leading to strong, narrow-band GW emission, as displayed in Fig.\ref{fig:3b} 
\cite{2007CoPhC.177..288C,2007PhRvL..98z1101O,
2008A&A...490..231S}.
At present little is known about the true nature 
of the so-called low $T/|W|$ instability.
Previous work has so far failed to establish
(for example) an analytical instability criterion, as
was pointed out by \cite{2009CQGra..26f3001O}.
We addressed two relevant 
questions regarding the 
so-called `low $T/|W|$' instability in the context 
of stellar core collapse: i) Which is the  
minimum $\beta$ value required in self-consistent
core-collapse simulation to trigger 
the onset of the instability?
This is important to know,
since 
most stars which undergo a core collapse
rotate only slowly \cite{2005ApJ...626..350H}; furthermore,
it was pointed out by \cite{2008PhRvD..78f4056D}
that even fast rotating PNS can never 
accrete enough angular momentum to reach the 
$\beta_{dyn}$ value required for the onset of the
classical bar mode instability. 
ii) How does the inclusion of deleptonization in 
the postbounce phase quantitatively alter the 
GW signal?
So far, 3D models have not included 
spectral neutrino physics in the postbounce phase. 
To study i), we systematically change the rotation rate
while keeping the other model parameters fixed.
The minimum $T/|W|$ value we found in our parameter range
to trigger the instability was $\beta_{b}\sim2.3\%$ at core bounce
(model R3), which is considerably lower than seen 
in previous studies (\cite{2007PhRvL..98z1101O} found
$\beta_{b}\sim9\%$, while \cite{2008A&A...490..231S}
found $\beta_{b}\sim5\%$). 
Furthermore, we find that centrifugal forces set a limit to 
the maximum frequency of the GW signal around $\sim$ 900Hz. 
The faster the initial 
rotation rate, the stronger the influence
of centrifugal forces, which slow down the postbounce advection
of angular momentum onto the PNS. The result is a slower 
rotation rate, a lower pattern speed and thus GW emission
at lower frequencies (see Tab.\ref{tabone}).
In order to address ii), we carried out `leakage' model R5$_{L}$. 
This model shows 5-10 $\times$ larger 
maximum amplitudes due to the nonaxisymmetric dynamics compared
to its hydrodynamical counterpart R5 that neglects neutrino 
cooling (see Fig. \ref{fig:3b}). 
This suggests that the treatment of postbounce neutrino cooling
plays an important role when it comes to the quantitative prediction
of GW signals from a low $\beta$ instability.
The neutrino cooling during the postbounce phase leads
to a more compact PNS with a shorter dynamical timescale
compared to the purely hydrodynamical treatment. 
This in turn is reflected in the dynamical evolution.
The shock wave stalls at considerably smaller radii and becomes
more quickly unstable to azimuthal fluid modes. Since there is much 
more matter in the unstable region of this model, the unstable
modes grow faster, causing the emission of much more powerful 
GWs.
However, we again point out that our leakage scheme 
overestimates the compactification of the PNS due to neutrino cooling.
The `reality' for the strength of GW emission therefore
should lay in between the results from the 
pure hydrodynamical- and leakage treatment.

\section{Summary and outlook}

We have presented the GW signature of
eleven 3D core-collapse simulations
with respect to variations in the spatial grid
resolution, the underlying 
neutrino transport physics and the initial rotation rate.
Our results show that in case of 
non- and slowly rotating models 
the GWs emitted during the first 
20ms after bounce are predominantly 
due to entropy driven `prompt' 
convection.
It turns out that the crucial parameter to 
study this stochastic phenomenon 
is the choice of the
spatial resolution and not the 
inclusion of a neutrino transport scheme. 
This parameter has a twofold
effect:
Firstly, it governs the influence of 
numerical noise, since a 
better resolution leads to lower
numerical seed perturbations and 
thus smaller grid alignment effects. 
Therefore, the GW amplitude right at core
bounce is smaller for higher spatial resolution.
Secondly, it enhances the ability
to follow dynamical features, as
better numerical resolution causes 
less numerical dissipation in the system,
which eases the dynamical effects which follow, 
leading to larger GW amplitudes after the 
core bounce compared to less resolved models.
The lepton driven convection is the central engine for the 
later dynamical postbounce evolution of the PNS 
($t\gtrsim 20$ms) and hence 
the GW emission.
Our findings and comparisons with 
state of the art 2D simulations of 
\cite{2009A&A...496..475M} suggest that 
the radial location and size 
of the convectively 
unstable layers are 
the key controlling factor for the outcome 
of the GW prediction, as they define 
the timescale and the dynamical behaviour 
of the overturning matter.
Here we find a large sensitivity to
the numerical approach of the neutrino transport 
scheme.
In our rotational core-collapse simulations, 
nonaxisymmetric dynamics develops for
models with a rotation rate of 
$\beta_{b} \gtrsim 2.3\%$ at core bounce.
Beyond this value, which is considerably lower than
found in previous studies (e.g. \cite{2007PhRvL..98z1101O,
2008A&A...490..231S}),
all models 
become subject to the 'low $T/|W|$'
instability of dominant m=1 or m=2 character 
within several ms after bounce (Fig.\ref{fig:3a}).
The fact that the effectively measured GW amplitude
scales with the number of GW cycles $N$ as 
$h_{eff}\propto h\sqrt{N}$ \cite{1989thyg.book.....H} 
suggests that the detection
of such a signal is tremendously enhanced. 
Moreover, we point out that the inclusion
of deleptonization during the postbounce phase
causes a compactification of the PNS which 
enhances the absolute values of the GW amplitudes 
up to a factor of ten with respect to a lepton-conserving
treatment.

The major limitation of our code now is
in the monopole treatment of gravity, since it 
cannot account for spiral structures, which could be reflected
in GW. 
We are currently working on the improvement of this issue.
Furthermore, the IDSA includes at present only the dominant 
reactions relevant to the neutrino transport problem 
(see \cite{2009ApJ...698.1174L} for details).
Future upgrades will also include contributions from 
electron-neutrino scattering, which are indispensable
during the collapse phase. The inclusion of this reaction 
will also make the cumbersome switch of the parametrization to
the IDSA at bounce obsolete.
Finally, we work on the inclusion of $\mu$ and $\tau$
neutrinos, which are very important for the cooling of the PNS
to its final stage as neutron star.

\ack
The authors would like to thank C. D. Ott for stimulating and 
useful discussions and F.-K. Thielemann for his
support.
This work was supported by a grant from the Swiss National 
Supercomputing Centre-CSCS under project ID s168.
We acknowledge support by the Swiss National Science Foundation under grant
No. 200020-122287 and PP00P2-124879.
Moreover, this work was supported by CompStar, 
a Research Networking Programme of the European Science Foundation.
Further thanks go to 
John Biddiscombe and Sadaf Alam from the Swiss Supercomputing Centre CSCS
for the smooth and enjoyable collaboration.


\end{document}